# Balancing spatial and non-spatial variation in varying coefficient modeling: a remedy for spurious correlation

Daisuke Murakami[1,*], Daniel A. Griffith[2]

[1]Department of Data Science, Institute of Statistical Mathematics,

10-3 Midori-cho, Tachikawa, Tokyo, 190-8562, Japan

Email: dmuraka@ism.ac.jp

[2]School of Economic, Political and Policy Science, The University of Texas, Dallas,

800 W Campbell Rd, Richardson, TX, 75080, USA

Email: dagriffith@utdallas.edu

* Corresponding author

Abstract

This study discusses the importance of balancing spatial and non-spatial variation in spatial regression modeling. Unlike spatially varying coefficients (SVC) modeling, which is popular in spatial statistics, non-spatially varying coefficients (NVC) modeling has largely been unexplored in spatial fields. Nevertheless, as we will explain, consideration of non-spatial variation is needed not only to improve model accuracy but also to reduce spurious correlation among varying coefficients, which is a major problem in SVC modeling. We consider a Moran eigenvector approach modeling spatially and non-spatially varying coefficients (S&NVC). A Monte Carlo simulation experiment comparing our S&NVC model with existing SVC models suggests both modeling accuracy and computational efficiency for our approach. Beyond that, somewhat surprisingly, our approach identifies true and spurious correlations among coefficients nearly perfectly, even when usual SVC models suffer from severe spurious correlations. It implies that S&NVC model should be used even when the analysis purpose is modeling SVCs. Finally, our S&NVC model is employed to analyze a residential land price dataset. Its results suggest existence of both spatial and non-spatial variation in regression coefficients in practice. The S&NVC model is now implemented in the R package spmoran.

# Introduction

Regression problems in the presence of spatial dependence and heterogeneity, which are common properties of spatial data (see Anselin 2010), have been studied in geostatistics (e.g., Cressie and Wikle 2011), spatial econometrics (LeSage and Pace 2009), spatial statistics (Cliff and Ord 1972), and other applied fields. Modeling of spatial varying coefficients (SVC) is a major recent concern in these fields (see Fotheringham et al. 2003; Wheeler and Páez 2010). SVC modeling estimates or predicts $NK$ coefficients using $N$ samples, where $K$ is the number of covariates; thus, a SVC model is essentially unidentifiable (Wheeler and Tiefelsdorf 2005). Assumptions are introduced to identify them. For example, geographically weighted regression (GWR; Brunsdon et al. 1996; Fotheringham et al. 2003) assumes greater weights on nearby samples to estimate local regression coefficients. Bayesian SVC models (Gelfand et al. 2003; Finley et al. 2009; Wheeler and Waller 2009) and Moran eigenvector-based SVC models (Griffith 2008; Murakami et al. 2017) assume spatially dependent map patterns underlie regression coefficients.

In (non-spatial) applied statistics, approaches for modeling coefficients smoothly varying with respect to one or more covariates, which we label non-spatially varying coefficients (NVC), have been developed (e.g., Hastie and Tibshirani 1993; Fan and Zhang 2008; Wang and Xia 2009; Hu and Xia 2012). For example, the unemployment rate may influence the crime rate only if it exceeds a threshold. An NVC that varies according to the unemployment rate would be useful to model such influence. NVC modeling has been applied to time-series analysis (e.g. Dangl and Halling 2012), quantile regression analysis (e.g. Wang et al. 2009), and meta-analysis (e.g., Bonett 2010), among other approaches. As with SVC modeling, a local approach exists for estimating coefficients using nearby samples in a feature space (see Park et al. 2015 for review), and a global approach exists for fitting a smooth function to modeling NVCs. Today, the additive model, which includes the global NVC model as a special case, is widely used to predict varying effects, group effects, and other effects with computational efficiency (e.g., Umlauf et al. 2015; Lindgren and Rue 2015; Wood et al. 2017). Kneib et al. (2009),

Franco-Villoria et al. (2019), among others, have applied additive models to predict SVCs.

Past studies have customarily assumed only SVC or NVC per covariate.[1] However, it is more reasonable to assume both for each covariate for the three reasons explained below. First, it is important to appropriately model the phenomenon under observation. For example, in the case of a regression analysis estimating the influence of crime ratio on land prices, the influence may (a) vary over space (e.g., crime ratio might decrease land price only in a certain area such as the city center) or (b) vary depending on the crime ratio (e.g., there might only be an impact if the ratio exceeds a threshold). SVCs are required to consider (a) while NVCs are needed to consider (b). Because (a) and (b) can co-occur, SVC and NVC are needed to be considered simultaneously.

Second, balancing SVC and NVC is necessary to enhance model accuracy while reducing model complexity. The two have considerably different complexity; SVCs, which are defined in 2- or 3-dimensional geographical space constrained by oceans, mountains, cities, and many other physical and cultural objects, are more complex than NVCs, which are typically defined in a 1-dimensional feature space. Misuse of SVC despite the true process being NVC, makes the model too complex and destabilizes the model estimation. By contrast, misuse of NVC in the presence of SVC as the true process renders the model too simple and reduces its accuracy.

Third, consideration of SVC and NVC can reduce multicollinearity among varying coefficients, which can lead to spurious correlation among these coefficients and decrease modeling accuracy. If estimated varying coefficients are correlated while the true coefficients are not, the correlation is called spurious correlation in this study (see the "Motivative example" section). Multicollinearity is recognized as a major problem in SVC modeling (Wheeler and Tiefelsdorf 2005; Páez et al. 2011; Fotheringham and Oshan 2016). A wide variety of studies have attempted to reduce multicollinearity among SVCs through regularization or other approaches (e.g., Wheeler 2007, 2009; Bárcena et al. 2014; Franco-Villoria et al. 2018; Comber et al. 2018; Griffith et al. 2019). Yet, there is still room for further consideration for more efficient remedy.

SVCs are collinear because they implicitly share the same basic functions whereas those for NVCs are independently defined for each covariate as we will explain later. A reasonable way to reduce the multicollinearity is to define the coefficients by the sum of SVC and NVC. The resulting S&NVC includes SVC as an inflator of collinearity and NVC as a deflator of collinearity. Although the S&NVC model is a relatively minor update of the SVC model, it is worth exploring because balancing SVC and NVC could reduce multicollinearity among varying coefficients, which is one of the biggest problems in (spatially) varying coefficient modeling (see Wheeler and Tiefelsdorf 2005).

This study aims to examines the usefulness of S&NVC modeling. The next section introduces a motivative example. The section, "A Monte Carlo simulation experiment," compares our S&NVC model with SVC models to explore the accuracy and stability of our approach, while the section titled "An empirical application" applies our approach to a residential land price dataset. Finally, our discussion is closed by the "Concluding remarks" section.

## Motivative example

This section illustrates the importance of considering NVC in spatial analysis through a toy example. We generate synthetic data on 40 by 40 grids the $X$ and $Y$ coordinates of which are defined by $px \in \{1, \dots, 40\}$ and $py \in \{1, \dots, 40\}$ respectively. The explained variable on the $i$-th grid is generated from

$$y_i = x_{i,1}\beta_{i,1} + x_{i,2}\beta_{i,2} + \varepsilon_i, \qquad \varepsilon_i \sim N(0, 0.2^2), \tag{1}$$

where $x_{i,1}$ is a covariate defined by the distance from the center coordinates of $(px, py) = (20, 20)$. We assume a likely situation that the influence of $x_{i,1}$ decays exponentially following $\beta_{i,1} = \exp\left(-\frac{x_{i,1}}{20}\right)$. For example, the distance from an urban park may have a distance-decaying impact on residential land prices. Likewise, $x_{i,2}$ is defined by the distance from $(px, py) = (1, 1)$ and $\beta_{i,2} = \exp\left(-\frac{x_{i,2}}{40}\right)$, which has a larger-scale map pattern, than $\beta_{i,1}$.

Although $\beta_{i,1}$ and $\beta_{i,2}$, which vary depending on $x_{i,1}$ and $x_{i,2}$ respectively, are NVCs according to our definition, they have smooth spatial patterns, as shown in Figure 1 (left). It seems acceptable to estimate or predict $\beta_{i,1}$ by fitting an SVC model, as is customarily done in spatial statistics. To verify this, we predict $\beta_{i,1}$ by fitting a Moran coefficient-based SVC model (MSVC model) and a NVC model, respectively. The Methodology section explains the details of the MSVC and NVC models. We also fit the GWR model (without intercept following Eq. 1) with the exponential kernel (GWR) and the GWR with the exponential adaptive kernel (GWRa), which tends to be more stable than GWR (Fotheringham et al., 2003).

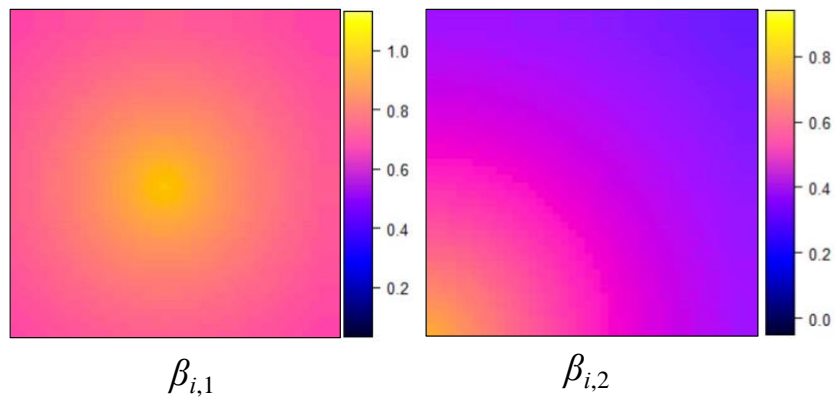

Figure 1: True varying coefficients

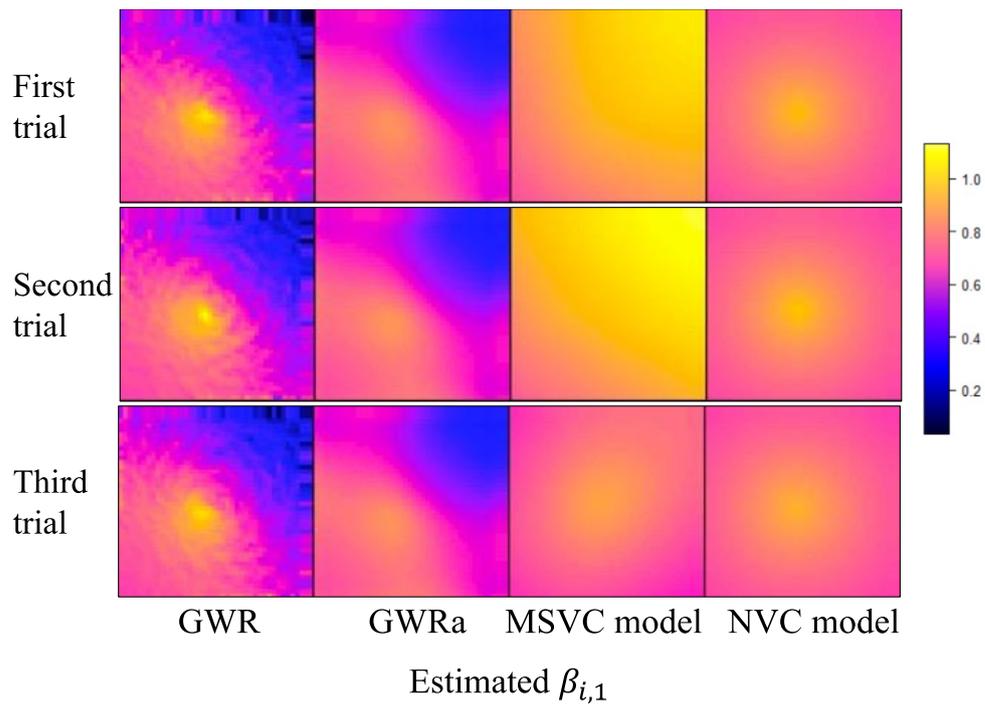

Estimated $\beta_{i,1}$

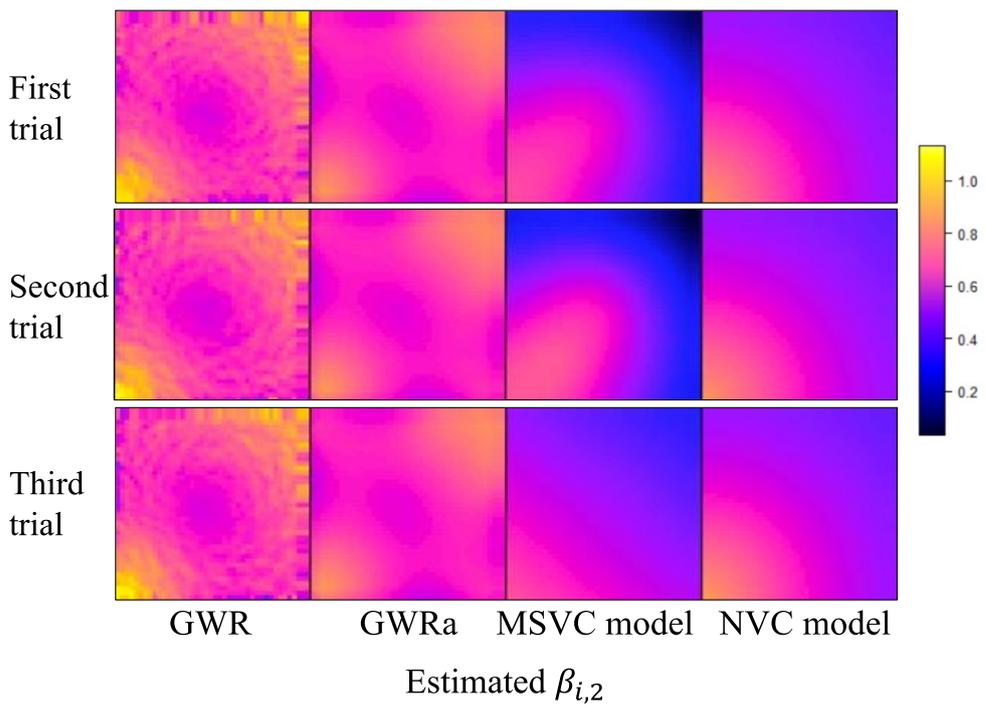

Estimated $\beta_{i,2}$

Figure 2: Varying coefficients obtained in the three trials. The SVC models (GWR, GWRa, the MSVC model) assume that the coefficients vary smoothly over space, while the NVC model assumes the coefficients vary depending on the covariates.

These coefficients were predicted three times while varying the $\varepsilon_i$ values. Figure 2 displays the SVC and NVC values obtained from each model. Unfortunately, all the SVC models (GWR, GWRa, and the MSVC model) failed to identify the true pattern. Although it is not shown here, the same was true for the multiscale GWR (e.g., Fotheringham et al., 2017). This indicates that the SVC model may perform poorly if coefficients vary depending on the covariate.

The mean correlation coefficients between the estimated $\beta_{i,1}$ and $\beta_{i,2}$ values are -0.498 for GWR, -0.590 for GWRa, and -0.639 for the MSVC model. The absolute values are considerably larger than the true correlation coefficient value of 0.088. In other words, the SVC models suffer from spurious correlation. As explained in the "Properties of the model" section, this spurious correlation is attributable to the variance inflation induced by wrongly estimating NVCs by fitting SVCs, which are collinear and assume much more complex patterns than NVC.

By contrast, the coefficients predicted from the NVC model are accurate across the cases. The correlation coefficient between the estimated varying coefficients is 0.102, which is close to the true value (0.088). The toy example suggests that, although NVC has rarely been used in spatial statistics, it is potentially helpful for stably modelling (spatially) varying coefficients. Still, NVC can only describe patterns depending on covariates; SVC is needed to describe complex spatial patterns. The next section introduces an approach for stably and accurately modeling varying coefficients using SVC and NVC.

## Spatially and non-spatially varying coefficient (S&NVC) modeling

This section explains our S&NVC modeling approach. The first subsection introduces our model. The next subsection explains the properties of our model focusing on stability and indefinability. The subsequent section explains how to estimate our model, and the last subsection analyzes its computational efficiency.

### The Model

This study considers the following S&NVC model:

$$\mathbf{y} = \sum_{k=1}^{K} \mathbf{x}_k \circ \boldsymbol{\beta}_k + \boldsymbol{\varepsilon}, \qquad \boldsymbol{\beta}_k = b_k \mathbf{1} + \boldsymbol{\beta}_k^{(s)} + \boldsymbol{\beta}_k^{(n)}, \qquad \boldsymbol{\varepsilon} \sim N(\mathbf{0}, \sigma^2 \mathbf{I}), \qquad (2)$$

where $\mathbf{y}$ is a vector of response variables which are observed at $N$ sample sites distributed in a 2-dimentional Euclidean space. $\mathbf{x}_k$ is a vector of the $k$-th covariate, $\mathbf{0}$ is a vector of zeros, $\mathbf{I}$ is an identity matrix, and " $\circ$ " is the operator multiplying each element of the vector in the left-hand side with each element of the matrix in the right-hand side. The coefficient vector $\boldsymbol{\beta}_k$ is defined by [mean: $b_k\mathbf{1}$] + [SVC: $\boldsymbol{\beta}_k^{(s)}$] + [NVC: $\boldsymbol{\beta}_k^{(n)}$], where $b_k$ is a parameter, and $\mathbf{1}$ is a vector of ones. For identifiability, we assume zero means for the SVC and NVC. Eq. (2).

In geostatistics, SVCs have been modeled using mean-centered Gaussian processes (GP; e.g., Gelfand et al., 2003) the covariance of which is specified by a distance-decaying function modeling spatial dependence. The GP-based SVC model is stable and accurate (Wheeler and Calder, 2007; Wheeler and Waller, 2009). However, without approximation, the estimation can be very slow even for moderate samples (e.g., $N > 2{,}000$) because of the need to iteratively evaluate the inverse of the covariance matrix for each SVC (see Finley et al., 2009, 2012).

For fast computation, we consider the following positively dependent spatial process, which is derived from the mean-centered GP (see Appendix.1):

$$\boldsymbol{\beta}_k^{(s)} = \mathbf{E}^{(s)}\boldsymbol{\gamma}_k^{(s)}, \qquad \boldsymbol{\gamma}_k^{(s)} \sim N\big(\mathbf{0}, \tau_{k(s)}^2\boldsymbol{\Lambda}^{\alpha_k}\big). \tag{3}$$

$\mathbf{E}^{(s)}$ is a matrix that consists of the $L$ eigenvectors corresponding positive eigenvalue of $\mathbf{MCM}$, which is a doubly-centered proximity matrix in which $\mathbf{M} = \mathbf{I} - \mathbf{1}\mathbf{1}'/N$ is the centering operator and $\mathbf{C}$ is a proximity matrix with zero diagonals and off-diagonals specified by a distance-decaying function. $\boldsymbol{\Lambda}$ is a diagonal matrix the elements of which are the positive eigenvalues. As detailed in Appendix.1, Eq.(3), which is called the MSVC model in the "Motivative example" section, is derived to capture all the positively dependent variation, which is interpretable in terms of the Moran coefficient, in the GP. Because GWR and other SVC models assume positively dependent coefficients whose values are smoothly varying over space, the assumption of positive spatial dependence is reasonable. Following Dray et al. (2006), the $(i, j)$-th element of the $\mathbf{C}$ matrix is specified by $c_{ij} = \exp(-d_{ij}/r)$, where $d_{ij}$ is the Euclidean distance between the sample sites $i$ and $j$, and $r$ is the maximum distance in the minimum spanning tree connecting the sample sites.

$\tau^2_{k(s)}$ is the parameter estimating the variance of the SVC. In the absence of positive spatial dependence in $\boldsymbol{\beta}_k$, $\tau^2_{k(s)}$ becomes zero and the SVCs take a uniform value. This parameter is helpful for preventing overfitting. $\alpha_k$ is the parameter estimating the Moran coefficient (MC) value or spatial scale of the SVC; as $\alpha_k$ grows, the expectation of the MC value increases and asymptotically converges to the maximum possible value (See Figure 3).

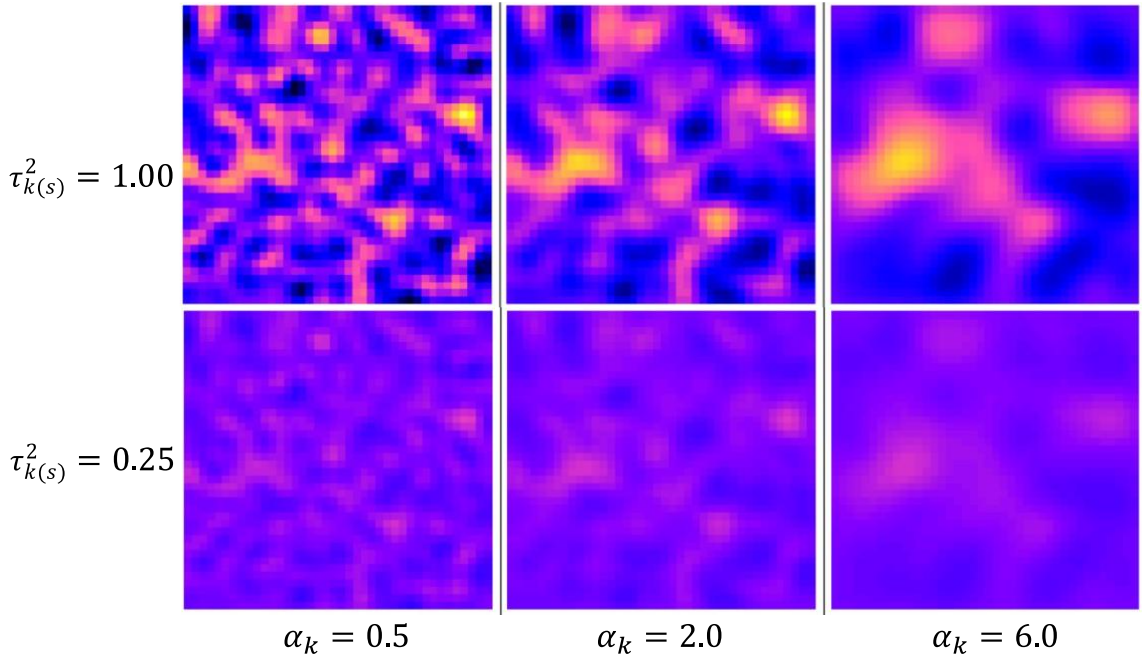

Figure 3: SVCs generated from Eq. (3) on 40 by 40 grids. The same random seed is used for each generation. As illustrated in this figure, $\alpha_k$ controls the spatial scale of the process, while $\tau^2_{k(s)}$ controls the variance.

For large samples (e.g., $N > 8,000$), a fast approximation is available for the eigen-decomposition (Murakami and Griffith, 2019a). After the exact or approximate eigen-decomposition is performed, the model is estimated computationally efficiently, as explained in Murakami and Griffith (2019b; 2020). Murakami et al. (2017) and Murakami and Griffith (2019) confirmed the accuracy of the eigenvector-based SVC modeling with and without approximation. We use the eigenvector-based approach modeling SVCs accurately, computationally efficiently, and in an interpretable manner in terms of the Moran coefficient.

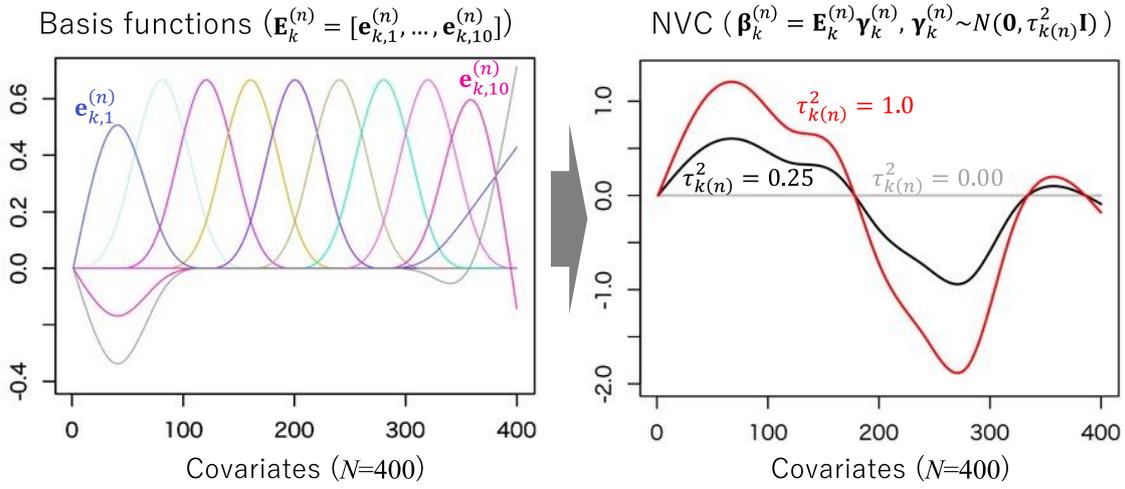

Figure 4: Example of basis functions ($L_k^{(n)} = 10$) generated from the covariate whose value ranges between 0 and 400 (left) and NVC (right). Natural splines are used for the basis functions. $\boldsymbol{\gamma}_k^{(n)}$ is generated through random sampling from $N\big(\mathbf{0}, \tau_{k(n)}^2 \mathbf{I}\big)$. As illustrated, the $\tau_{k(n)}^2$ parameter regularizes the variance of the NVC. If $\tau_{k(n)}^2 = 1.0$, the predicted NVC suggests strong positive impact for covariates whose values are around 100 while negative impact whose values are roughly between 200 and 300.

Meanwhile, the NVC $\boldsymbol{\beta}_k^{(n)}$ is given as follows:

$$\boldsymbol{\beta}_k^{(n)} = \mathbf{E}_k^{(n)}\boldsymbol{\gamma}_k^{(n)}, \qquad \boldsymbol{\gamma}_k^{(n)} \sim N\left(\mathbf{0}, \tau_{k(n)}^2\mathbf{I}\right), \tag{4}$$

where $\mathbf{E}_k^{(n)}$ is a matrix that consists of $L_k^{(n)}$ basis functions for the $k$-th covariate. The basis functions are defined by spline functions generated from $\mathbf{x}_k$, and $\tau_{k(n)}^2$ is a variance parameter. $10 - 20$ of basis functions are typically sufficient for modeling (1-dimensional) NVC. Eq. (4) is widely used for modeling non-linear effects (with respect to $\mathbf{x}_k$) in additive modeling studies (e.g., Ruppert et al. 2003; Wood, 2017). See Figure 4 for an intuitive image of NVC.

Properties of the model

By substituting Eq. (3) and Eq.(4) into Eq.(2), our model can be written as

$$\mathbf{y} = \mathbf{Z}\mathbf{B} + \boldsymbol{\varepsilon}, \qquad \boldsymbol{\varepsilon} \sim N(\mathbf{0}, \sigma^2\mathbf{I}), \tag{5}$$

where $\mathbf{B} = \left[b_1, \dots, b_K, \boldsymbol{\gamma}_1^{(s)\prime}, \dots, \boldsymbol{\gamma}_K^{(s)\prime}, \boldsymbol{\gamma}_1^{(n)\prime}, \dots, \boldsymbol{\gamma}_K^{(n)\prime}\right]\prime$ is a vector that stacks all the fixed and random coefficients. $\mathbf{Z}$ is the design matrix that is given as

$$\mathbf{Z} = \left[\mathbf{x}_1, \dots, \mathbf{x}_K, (\mathbf{x}_1 \circ \mathbf{E}^{(s)}), \dots, (\mathbf{x}_K \circ \mathbf{E}^{(s)}), (\mathbf{x}_1 \circ \mathbf{E}_1^{(n)}), \dots, (\mathbf{x}_K \circ \mathbf{E}_K^{(n)})\right], \tag{6}$$

Our model yields a simple linear regression model if the random coefficients $\{\boldsymbol{\gamma}_k^{(s)}, \boldsymbol{\gamma}_k^{(n)}\}$ in $\mathbf{B}$ are replaced with fixed coefficients. Ordinary least squares (OLS) estimation is available for the fixed coefficients model. However, OLS-based SVC modeling performs poorly, as demonstrated by Helbich and Griffith (2016) and Murakami et al. (2017; 2019b). The low accuracy is attributed to the variance inflation due to the multicollinearity among the $LK$ spatial basis functions $(\mathbf{x}_1 \circ \mathbf{E}^{(s)}), \dots, (\mathbf{x}_K \circ \mathbf{E}^{(s)})$, which share the same $\mathbf{E}^{(s)}$ matrix; while the $L$ value, which is determined to model positively dependent variations, can be large, the large $L$ makes the multicollinearity even more severe.[2] The random coefficients $\{\boldsymbol{\gamma}_k^{(s)}, \boldsymbol{\gamma}_k^{(n)}\}$ play a critical role in reducing variance and improving the identifiability of the varying coefficients, as explained in the "Estimation" section.

The non-spatial basis functions $(\mathbf{x}_1 \circ \mathbf{E}_1^{(n)}), \dots, (\mathbf{x}_K \circ \mathbf{E}_K^{(n)})$ are less prone to be collinear because they do not share the same basis matrix. In addition, at most 10 to 20 basis functions are typically sufficient to model each NVC. If substitutable, it will be helpful to replace SVC with NVC to stably model varying coefficients. Figure 2 illustrates the usefulness of NVC. Still, SVC is more flexible than NVC in terms of spatial process modeling. It is reasonable to balance SVC, which is flexible but can be unstable, and NVC, which is stable but less flexible.

---

[2] To mitigate multicollinearity, it is reasonable to use only the eigenvectors explaining positive spatial dependence, which is dominant in many cases in regional science (Griffith, 2003). The use of a limited number of basis function is widely accepted in SVC modeling studies in geostatistics (e.g., Finley et al., 2009; 2012).

Estimation

Somewhat surprisingly, there is no general rigorous definition of the likelihood function for regression models including random effects (Bayarri and DeGroot, 1992; Lee and Nelder, 2005). However, a wide variety of generalized likelihoods have been proposed.

Among generalized likelihood approaches, the restricted likelihood maximization (REML) of Laird and Ware (1982), which is equivalent with the empirical Bayes estimation, is widely accepted for models with random effects (e.g., Wood, 2011; Bates, 2010; Hodges, 2013). [3] Suppose that $\boldsymbol{\gamma}_{1:K}^{(s)} \in \{\boldsymbol{\gamma}_1^{(s)}, \dots, \boldsymbol{\gamma}_K^{(s)}\}$ and $\boldsymbol{\gamma}_{1:K}^{(n)} \in \{\boldsymbol{\gamma}_1^{(n)}, \dots, \boldsymbol{\gamma}_K^{(n)}\}$, the restricted likelihood assumes $\{\mathbf{b}, \boldsymbol{\gamma}_{1:K}^{(s)}, \boldsymbol{\gamma}_{1:K}^{(n)}\}$ as nuisance and integrate out. The resulting our likelihood yields:

$$L(\boldsymbol{\theta}|\mathbf{y}) = \iiint L(\boldsymbol{\theta}, \mathbf{b}, \boldsymbol{\gamma}_{1:K}^{(s)}, \boldsymbol{\gamma}_{1:K}^{(n)}|\mathbf{y}) \, d\mathbf{b} d\boldsymbol{\gamma}_{1:K}^{(s)} d\boldsymbol{\gamma}_{1:K}^{(n)}, \tag{7}$$

---

[3] Alternatively, if we regard $\{\boldsymbol{\gamma}_k^{(s)}, \boldsymbol{\gamma}_k^{(n)}\}|k \in \{1, \dots, K\}$ as missing observations, a joint likelihood for the complete data $\{\mathbf{y}, \boldsymbol{\gamma}_k^{(s)}, \boldsymbol{\gamma}_k^{(n)}\}|k \in \{1, \dots, K\}$ can be defined. The likelihood is typically maximized by an expectation-maximization (EM)-type algorithm (e.g., Gelfand and Carlin, 1993). Unfortunately, this approach is unsuitable in our case because the number of elements in $\{\boldsymbol{\gamma}_k^{(s)}, \boldsymbol{\gamma}_k^{(n)}\}|k \in \{1, \dots, K\}$ can be very large, while the EM algorithm performs poorly in such a case. Besides, the joint likelihood maximization has a difficulty to clearly distinguish the distributions of random variables $\{\boldsymbol{\gamma}_k^{(s)}, \boldsymbol{\gamma}_k^{(n)}\}$ and observations $\mathbf{y}$ (Bayarri and DeGroot, 1992; Hodges, 2013). Because of these reasons, the joint likelihood is rarely used for additive mixed modeling.

where

$$L(\boldsymbol{\theta}, \mathbf{b}, \boldsymbol{\gamma}_{1:K}^{(s)}, \boldsymbol{\gamma}_{1:K}^{(n)} | \mathbf{y}) = p(\mathbf{y} | \boldsymbol{\theta}, \mathbf{b}, \boldsymbol{\gamma}_{1:K}^{(s)}, \boldsymbol{\gamma}_{1:K}^{(n)}) \prod_{k=1}^{K} p(\boldsymbol{\gamma}_k^{(s)} | \boldsymbol{\theta}) p(\boldsymbol{\gamma}_k^{(n)} | \boldsymbol{\theta}). \qquad (8)$$

Eq. (8) penalizes the data likelihood $p(\mathbf{y} | \boldsymbol{\theta}, \mathbf{b}, \boldsymbol{\gamma}_{1:K}^{(s)}, \boldsymbol{\gamma}_{1:K}^{(n)})$, which is a likelihood given the random variables, by the SVC prior $p(\boldsymbol{\gamma}_k^{(s)} | \boldsymbol{\theta})$ and NVC prior $p(\boldsymbol{\gamma}_k^{(n)} | \boldsymbol{\theta})$ for the random coefficients. These three are obtained from Eqs. (2), (3), and (4), respectively. The variations explained by the data likelihood, SVC, and NVC priors are estimated by $\sigma^2$, $\tau_{k(s)}^2$, and $\tau_{k(n)}^2$, respectively. For example, if $\tau_{k(s)}^2 = \infty$ and $\tau_{k(n)}^2 = \infty$ for $k \in \{1, \ldots, K\}$, the priors have no impact, and the random coefficients estimator yield the OLS estimator, which is unbiased in terms of the data likelihood but suffers from variance inflation (see "Property of the model" section). By contrast, if $\tau_{k(s)}^2 = 0$ and $\tau_{k(n)}^2 = 0$, all the random coefficients have zero values, which are biased in terms of the data likelihood but free from variance inflation. The REML estimates the variance parameters to optimize the bias-variance trade-off measured by the restricted likelihood Eq. (7) in which the marginalization (i.e., integral) in this equation is required to eliminate bias in the variance parameter estimates. The resulting REML estimator, which is equivalent to the maximum-a-posteriori (MAP) estimator, yields the mode of the posterior distribution $p(\boldsymbol{\theta} | \mathbf{y})$. The mode is the most probable value given the data likelihood and the priors. The predictor for the S&NVC, which is unbiased and minimizes the expected predictive

error variance, is readily obtained after the variance parameter estimation. See Appendix 2 for further details about the REML.

The theoretical basis for the REML has been established (e.g., Laird and Ware, 1992; Wood, 2011; Marra and Wood, 2012). In particular, Reiss and Ogden (2009) provided a theory for finite samples that asserts that the variance parameter estimators are stable and tend to converge to the global optima. Wood (2011) confirmed this empirically. The REML is widely accepted as a method that stably estimates additive mixed models, including our model. For the likelihood maximization, we use the fast algorithm of Murakami and Griffith (2019b; 2020).

Computational efficiency

The S&NVC model estimation can be slow due to the $3K$ variance parameters $\{\boldsymbol{\theta}_1, \dots, \boldsymbol{\theta}_K\}$. For example, in a case with 5 covariates, our model needs to estimate 15 variance parameters whereas GWR has only one parameter that must be estimated numerically. Fortunately, the computational complexity of the fast REML scales well for both sample size and number of variance parameters; their approach took only 4,221 seconds to estimate a SVC model with 14 variance parameters and 10 million locations.[4]

---

[4] The estimation is done with parallel computation using R (version 3.6.2; https ://cran.r-proje

Note that the large number of parameters prohibits applying other ML/REML, the integrated nested Laplace approximation (INLA; Rue and Martino, 2007; Lindgren and Rue, 2015), which is increasingly popular in spatial statistics, or other estimation methods whose computational cost exponentially grows with respect to the number of variance parameters to be estimated. Use of the fast REML is reasonable in terms of stability and computational efficiency.

## A Monte Carlo simulation experiment

This section compares our approach with other spatial modeling approaches using a Monte Carlo experiment. After outlining our settings in the subsection below, estimation accuracy, robustness against spurious correlation, and computational efficiency are compared in the subsequent subsections, respectively.

### Outline

This section compares the basic linear regression model (LM), GWR, adaptive GWR (GWR$_A$), the MC-based SVC (SVC$_M$) model, and the proposed S&NVC (S&NVC$_M$) model. GWR estimates regression coefficients at the $i$-th site by assigning greater weights to nearby samples. The local weights are given by the exponential kernel $\exp\left(-d_{i,j}/r\right)$, where $r$ is optimized by minimizing a corrected Akaike Information Criterion (AICc). In GWR$_A$, the kernel is given by $\exp\left(-d_{i,j}/r_{i(m)}\right)$ in which $r_{i(m)}$ is the distance between the $i$-th site and the $m$-th nearest neighbor, where $m$ is optimized by AICc minimization. The adaptive GWR is more robust than the usual GWR (see, Fotheringham et al. 2003).

Modeling accuracy of these models is compared by fitting them to synthetic data generated by

$$\mathbf{y} = \boldsymbol{\beta}_1 + \mathbf{x}_2 \circ \boldsymbol{\beta}_2 + \mathbf{x}_3 \circ \boldsymbol{\beta}_3 + \boldsymbol{\varepsilon}, \qquad \boldsymbol{\varepsilon} \sim N(\mathbf{0}, 2^2\mathbf{I}). \tag{9}$$

The covariates are specified as

$$\mathbf{x}_k = \mathbf{1} + w_{s(x)}\left[\bar{\mathbf{C}}\mathbf{e}_{k(x)}\right] + \left(1 - w_{s(x)}\right)\left[\mathbf{u}_{k(x)}\right],$$

$$\mathbf{e}_{k(x)} \sim N(\mathbf{0}, \mathbf{I}), \qquad \mathbf{u}_{k(x)} \sim N(\mathbf{0}, \mathbf{I}), \tag{10}$$

where $[\bullet]$ denotes standardization of the vector $\bullet$. Spatial coordinates are two generated standard normal random variables. $\bar{\mathbf{C}}$ is the matrix that row-standardizes the $\mathbf{C}$ matrix, whose $(i, j)$-th element is $\exp\left(-d_{i,j}\right)$. Eq. (10) defines the covariate $\mathbf{x}_k$ by a sum of a spatially dependent process $\bar{\mathbf{C}}\mathbf{e}_{k(x)}$ and an independent process, whose shares are $w_{s(x)}$ and $1 - w_{s(x)}$, respectively. A larger $w_{s(x)}$ implies stronger spatially dependent variation in $\mathbf{x}_k$ and $\boldsymbol{\beta}_k$ that make modeling unstable (see Paciorek 2010).

The coefficients are specified as

$$\boldsymbol{\beta}_1 = \mathbf{1} + [\bar{\mathbf{C}}\mathbf{e}_1], \qquad \mathbf{e}_1 \sim N(\mathbf{0}, \mathbf{I}),$$

$$\boldsymbol{\beta}_2 = (0.5)\mathbf{1} + \tau_2[\bar{\mathbf{C}}\mathbf{e}_2], \qquad \mathbf{e}_2 \sim N(\mathbf{0}, \mathbf{I}),$$

$$\boldsymbol{\beta}_3 = (-2)\mathbf{1} + \tau_3[w_s[\bar{\mathbf{C}}\mathbf{e}_3] + (1 - w_s)[\mathbf{E}_3^{(n)}\mathbf{u}_3^{(n)}]], \tag{11}$$

$$\mathbf{e}_3 \sim N(\mathbf{0}, \mathbf{I}), \qquad \mathbf{u}_3^{(n)} \sim N(\mathbf{0}, \mathbf{I}),$$

The term $\mathbf{E}_3^{(n)}\mathbf{u}_3^{(n)}$ models non-spatial variation using $\mathbf{E}_3^{(n)}$, which is a matrix of 10 thin plate spline basis functions generated from $\mathbf{x}_3$. $\boldsymbol{\beta}_1$ is a spatially varying intercept filtering residual spatial dependence (see Tiefelsdorf and Griffith 2007; Chun et al. 2016), $\boldsymbol{\beta}_2$ is a SVC with variance $\tau_2^2$, and $\boldsymbol{\beta}_3$ is a S&NVC with variance $\tau_3^2$, with the share of SVC being $w_s$.

The simulation is iterated 200 times while varying $w_{s(x)} \in \{0.00, 0.40, 0.80\}$, $w_s \in \{0.00, 0.25, 0.50, 0.75, 1.00\}$, $(\tau_2^2, \tau_3^2) \in \{(1^2, 3^2), (3^2, 1^2)\}$, and $N \in \{50, 150, 1000\}$. These simulations are performed using R version 3.6.0 (https://cran.r-project.org/) on a 64 bit PC whose memory is 48 GB. We use a Mac Pro (3.5 GHz, 6-Core Intel Xeon E5 processor with 64 GB of memory). R (version 3.6.2; https ://cran.r-project.org/) is used for model estimation. The GWmodel package (Lu et al., 2014; https://cran.r-project.org/web/packages/GWmodel/index.html) is used to estimate GWR and GWR$_A$, and spmoran (Murakami 2017; https://cran.r-project.org/web/packages/spmoran/index.html) is used to estimate SVC$_M$.

Results: coefficient modeling accuracy

Prediction error for coefficients $\boldsymbol{\beta}_k$ are evaluated using root mean squared error (RMSE), which is defined as

$$RMSE[\boldsymbol{\beta}_k] = \sqrt{\frac{1}{200}\sum_{p=1}^{200}(\boldsymbol{\beta}_k - \widehat{\boldsymbol{\beta}}_{k(p)})'(\boldsymbol{\beta}_k - \widehat{\boldsymbol{\beta}}_{k(p)})}, \quad (12)$$

where $\widehat{\boldsymbol{\beta}}_{k(p)}$ is the predictor in the $p$-th of 200 iterations.

Figure 5 summarizes the RMSEs when $N = 1,000$ and $(\tau_2^2, \tau_3^2) = (1^2, 3^2)$, which represents weak SVC and strong S&NVC (i.e., a small $Var[\boldsymbol{\beta}_2]$ and large $Var[\boldsymbol{\beta}_3]$). As expected, the modeling error of SVC models, including GWR, GWR$_A$, and SVC$_M$, rapidly inflates as the share of spatial variation, $w_s$, decreases. The RMSE inflation gets severe when covariates have strong spatially dependent variation (i.e., $w_{s(x)} = 0.8$). These results confirm that SVC models are not appropriate in the presence of non-spatial variation in regression coefficients.

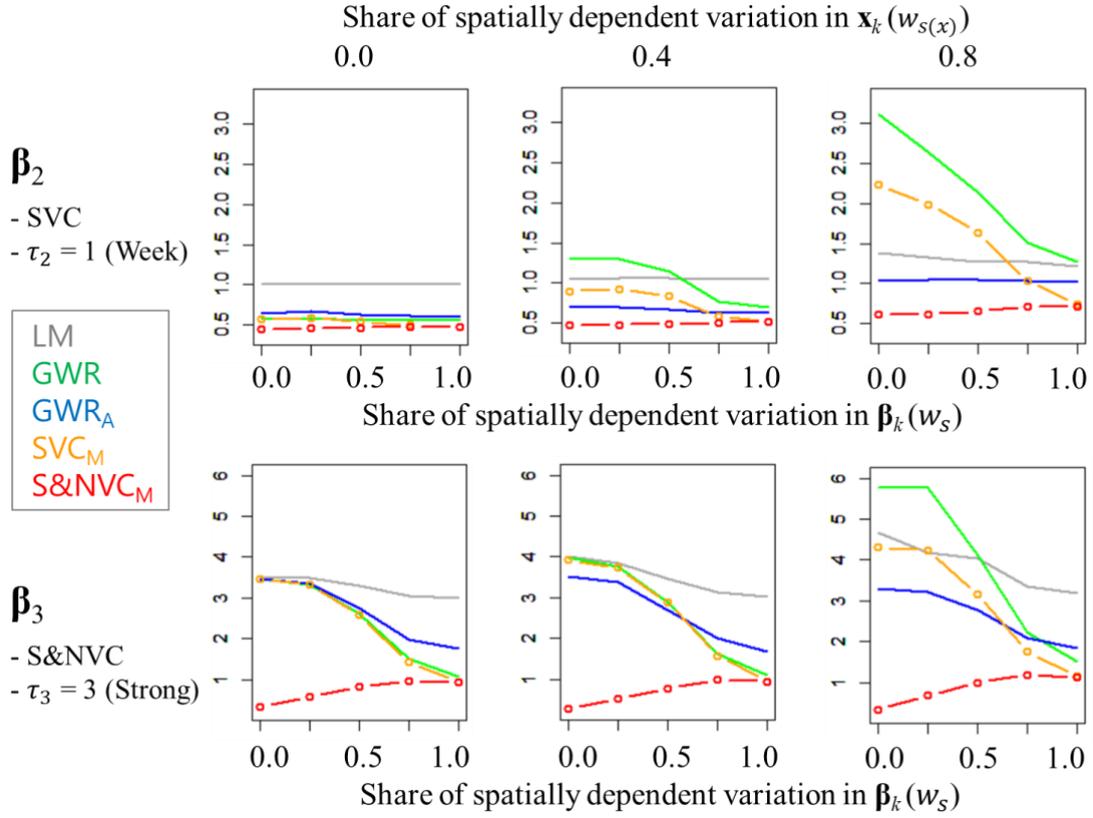

Figure 5: RMSEs for $\boldsymbol{\beta}_2$ and $\boldsymbol{\beta}_3$ when $(\tau_2^2, \tau_3^2) = (1^2, 3^2)$ and $N = 1{,}000$.

In contrast, the SVC models accurately predict $\boldsymbol{\beta}_2$, which is a pure SVC, when covariates are independently distributed. However, somewhat surprisingly, GWR and SVC$_M$ indicate greater $RMSE[\boldsymbol{\beta}_2]$ than LM in the presence of strong spatially dependent variation in covariates ($w_{s(x)} = 0.8$). In other words, SVC modeling is not appropriate if covariates are spatially dependent and more than one coefficient has strong non-spatial variation. Unfortunately, these conditions are likely in many real-world cases. In contrast, our S&NVC_M model accurately predict $\boldsymbol{\beta}_2$ and $\boldsymbol{\beta}_3$ across cases. The RMSE is substantially smaller than that for the SVC models when covariates are spatially dependent. Accuracy and stability of our S&NVC_M modeling approach is verified in the strong S&NVC cases.

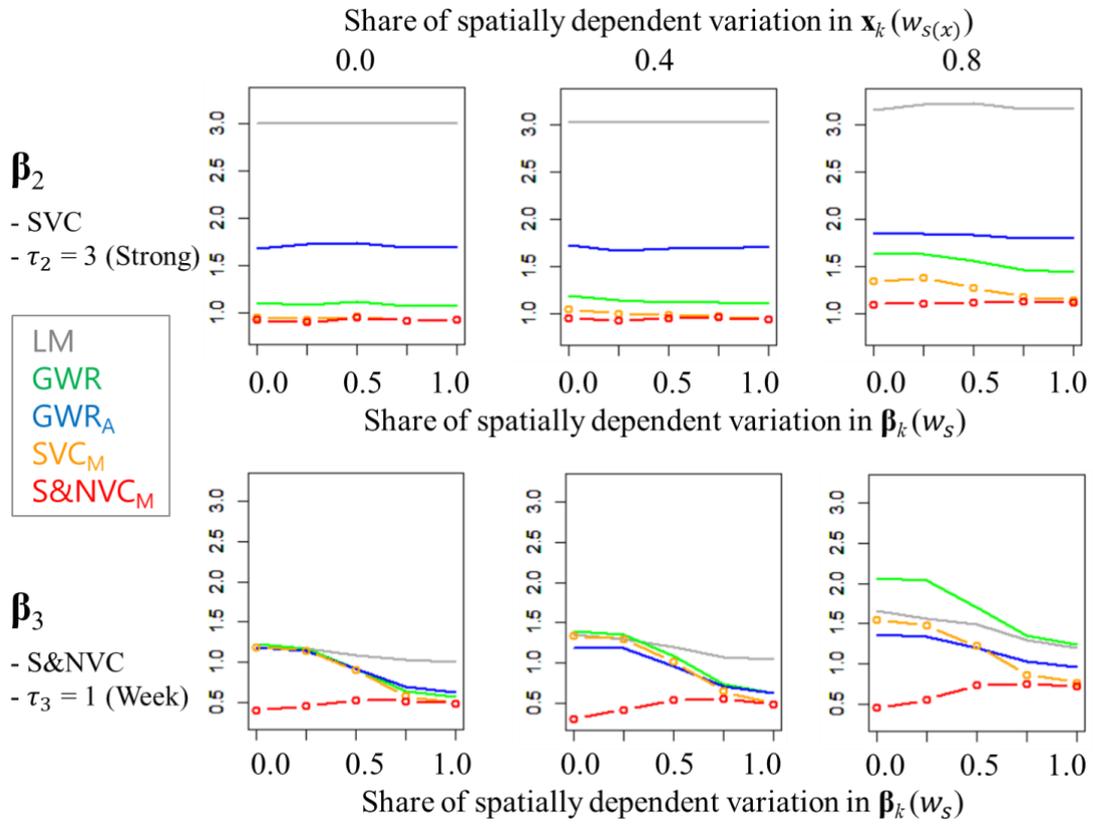

Figure 6: RMSEs for $\boldsymbol{\beta}_2$ and $\boldsymbol{\beta}_3$ when $(\tau_2^2, \tau_3^2) = (3^2, 1^2)$ and $N = 1{,}000$.

Figure 6 plots the RMSEs when $N = 1,000$ and $(\tau_2^2, \tau_3^2) = (3^2, 1^2)$, meaning strong SVC and weak S&NVC. Unlike the cases with strong S&NVC, the GWR, GWR$_A$, and SVC$_M$ each have a fairly small $RMSE[\boldsymbol{\beta}_2]$ even when $w_{s(x)} = 0.8$. SVC models are found to be acceptable in the absence of strong S&NVC. Still, S&NVC$_M$ tends to have a smaller $RMSE[\boldsymbol{\beta}_2]$ owing to the consideration of non-spatial variation in $\boldsymbol{\beta}_3$. Regarding $\boldsymbol{\beta}_3$, S&NVC$_M$ has considerably smaller RMSEs than its alternatives, as expected.

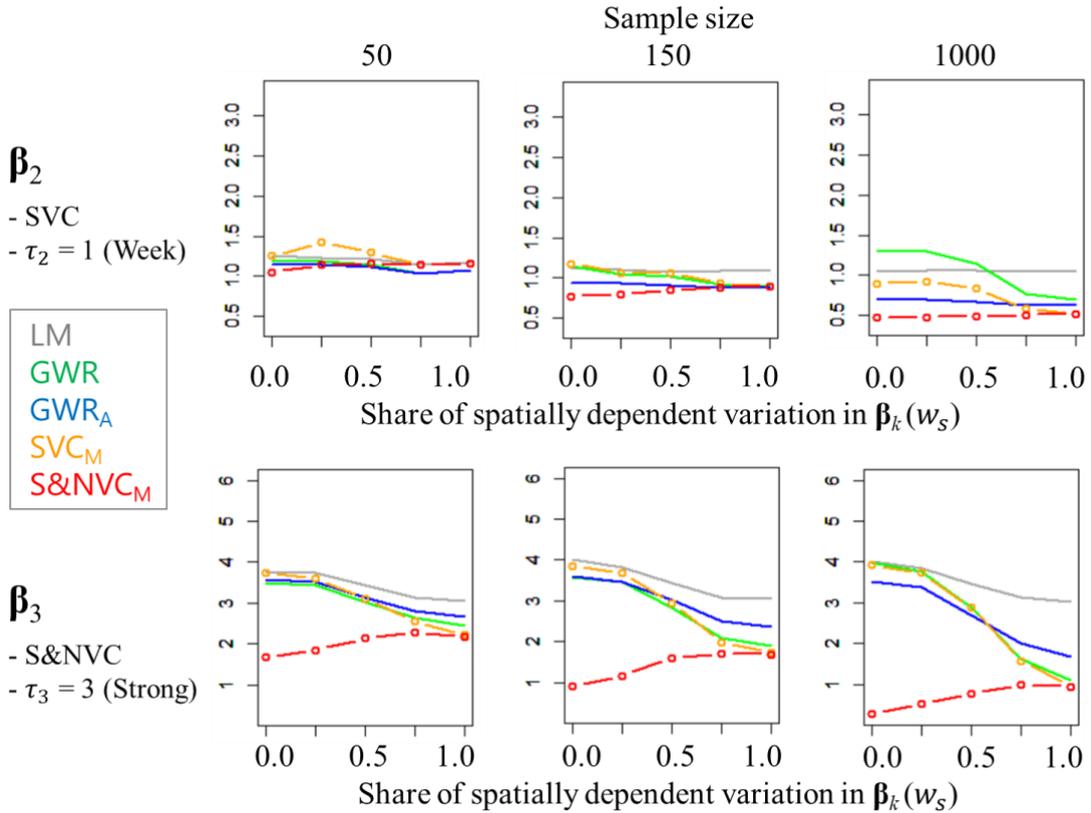

Figure 7: RMSEs for $\boldsymbol{\beta}_2$ and $\boldsymbol{\beta}_3$ when $N \in \{50, 150, 1,000\}$, $(\tau_2^2, \tau_3^2) = (1^2, 3^2)$, and $w_{s(x)} = 0.4$.

Figure 7 shows the RMSEs when $N \in \{50, 150, 1000\}$ and $(\tau_2^2, \tau_3^2) = (1^2, 3^2)$. This figure illustrates that our approach accurately predicts SVC and S&NVC even from small samples.

Figure 8 compares RMSE values for SVC$_M$ and S&NVC$_M$ across cases. This figure highlights that $RMSE[\boldsymbol{\beta}_k]$ of SVC$_M$ rapidly increases if both $\boldsymbol{\beta}_k$ and $\mathbf{x}_k$ have strong spatially dependent variation (e.g., $w_s = 1.0$ and $w_{s(x)} = 0.8$). This tendency is prominent in the presence of strong S&NVC ($\tau_3^2 = 3^2$). In contrast, S&NVC$_M$ accurately predicts coefficients across cases.

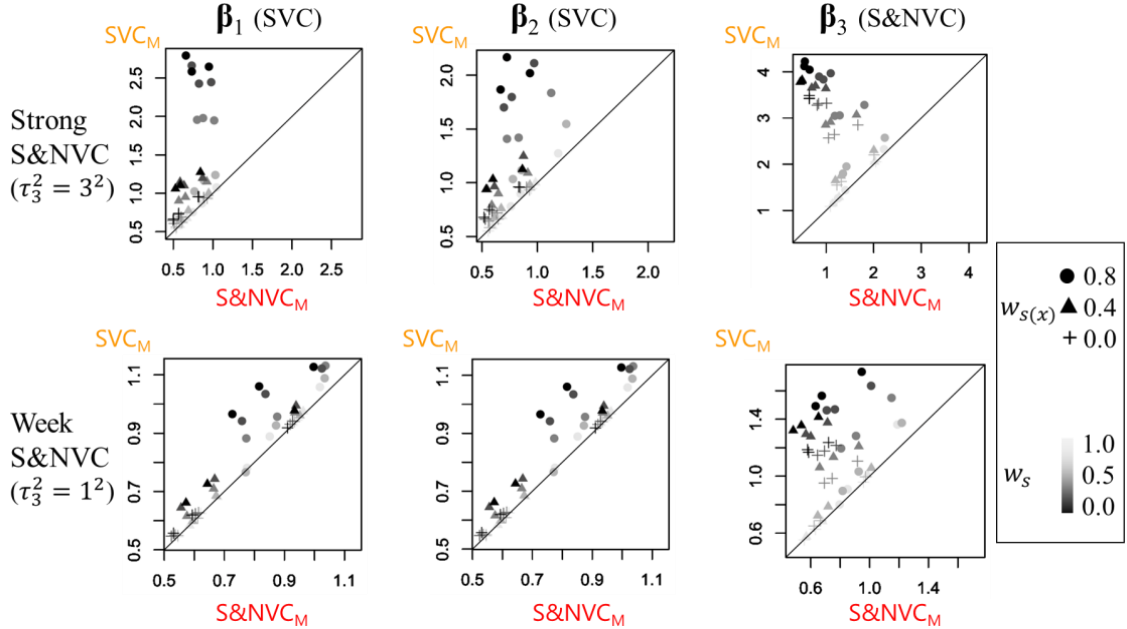

Figure 8: A comparison of RMSEs between SVC$_M$ and S&NVC$_M$ when $N = 1,000$.

Results: spurious correlation

This section summarizes robustness findings about our approach with respect to multicollinearity. Two types of correlation exist among S(N)VCs: (i) true correlation underlying S(N)VCs; and, (ii) spurious correlation attributable to collinearity between the spatial dependent patterns in S(N)VCs and covariates.

To reveal if our approach successfully identifies (i) true correlation while (ii) reducing spurious correlation, mean correlation coefficients (CCs) among true coefficients $\{\boldsymbol{\beta}_1, \boldsymbol{\beta}_2, \boldsymbol{\beta}_3\}$ and the CCs among predicted coefficients $\{\widehat{\boldsymbol{\beta}}_1, \widehat{\boldsymbol{\beta}}_2, \widehat{\boldsymbol{\beta}}_3\}$ are compared. Figure 9 compares CCs between true $\boldsymbol{\beta}_k$s with CCs between predicted $\widehat{\boldsymbol{\beta}}_k$s when $(\tau_2^2, \tau_3^2) = (1^2, 3^2)$, and Figure 10 summarizes the same kind of results when $(\tau_2^2, \tau_3^2) = (3^2, 1^2)$. SVC$_M$ predictors have greater CC values than the true CCs. In other words, SVC$_M$ suffers from spurious correlation. Although not shown here, GWR and GWR$_A$ estimates suffer from the same problem. These spurious correlations get severe as the share of spatial variation in S&NVC, $w_s$, and covariates, $w_{s(x)}$, increase. These results suggest that SVC models cannot avoid the spurious correlation problem.

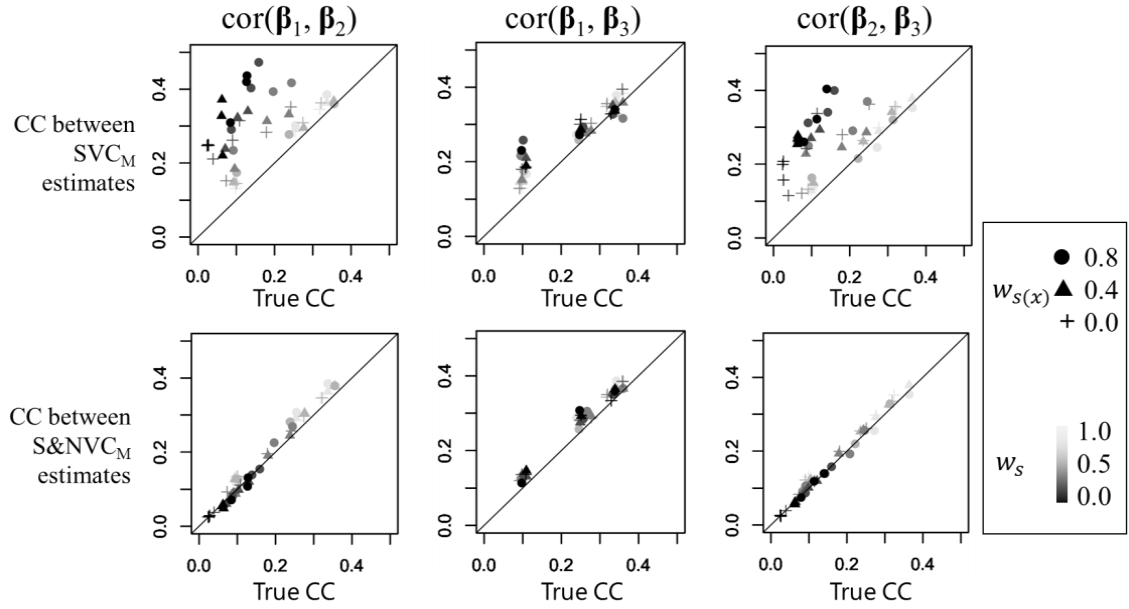

Figure 9: Correlation coefficients (CC) between $\boldsymbol{\beta}_k$s when $(\tau_2^2, \tau_3^2) = (1^2, 3^2)$. The *X*-axis represents CCs between true $\boldsymbol{\beta}_k$s, whereas the *Y*-axis represents CCs between estimated $\widehat{\boldsymbol{\beta}}_k$s.

CCs between S&NVC$_M$ predictors are almost the same as the true CCs. Based on the preceding result, our approach does not suffer from the spurious correlation problem. S&NVC modeling, which has been overlooked in spatial statistics, might actually be useful to reduce the spurious correlation problem in S(N)VC modeling.

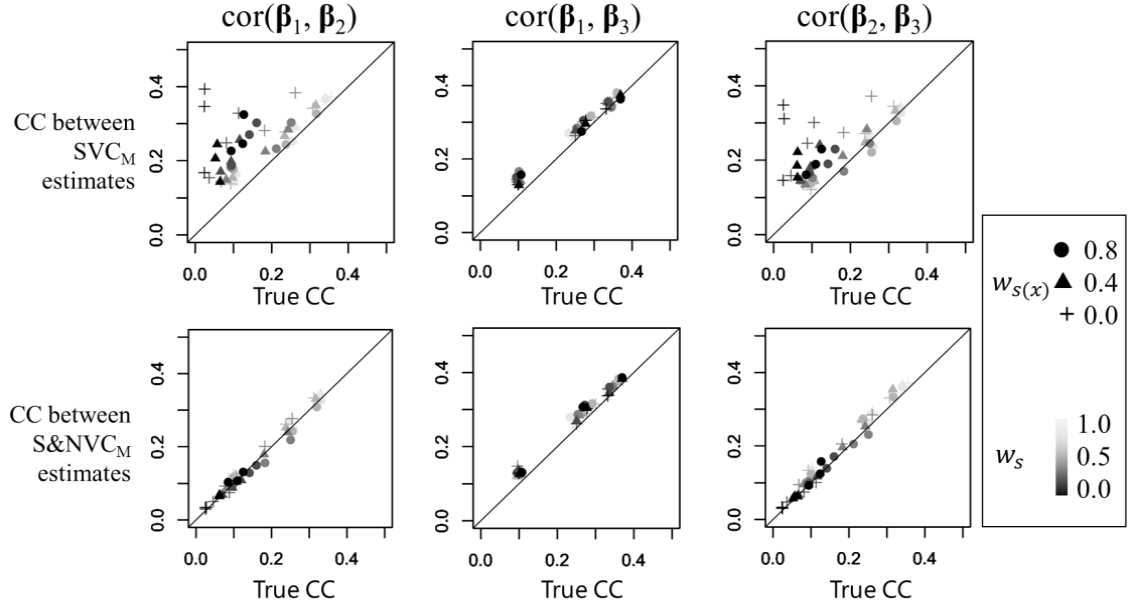

Figure 10: Correlation coefficients (CC) between $\boldsymbol{\beta}_k$s when $(\tau_2^2, \tau_3^2) = (3^2, 1^2)$. The *X*-axis represents CCs between true $\boldsymbol{\beta}_k$s, whereas the *Y*-axis represents CCs between estimated $\widehat{\boldsymbol{\beta}}_k$s.

Results: computation time

This section examines computational efficiency of S&NVC$_M$ by comparing its computation (CP) time with that for SVC$_M$, which is known to be computationally efficient (Murakami and Griffith, 2019b, 2020). The true data are generated with Eqs. (10) and (11), with $w_s = 0.5$, $w_{s(x)} = 0.4$, and $(\tau_2^2, \tau_3^2) = (1^2, 3^2)$. We estimate S&NVC$_M$ and SVC$_M$ models ten times for cases with $N \in \{1{,}000, 10{,}000, 50{,}000, 100{,}000, 150{,}000, 200{,}000\}$, recording the average CP times for estimation and Moran basis extraction for comparison purposes. Here, the 200 approximate Moran basis corresponding to the largest positive eigenvalues are used based on evidence that the 200 eigenvectors explain roughly 90% or more of positively dependent spatial variations if $N \leq 250,000$ (see Appendix 1 of Murakami and Griffith, 2019).

Figure 11 summarizes the average CP times. Owing to the fast REML, which compresses large matrices before estimation, the CP time increase of S&NVC$_M$ with respect to $N$ is similar to that for SVC$_M$, despite S&NVC$_M$ having $K$ additional variance parameters; S&NVC$_M$ also is available for large datasets. Note that, although we did not do so, REML can be parallelized for even larger samples (e.g., millions) for faster estimation while saving memory (see Murakami and Griffith, 2020).

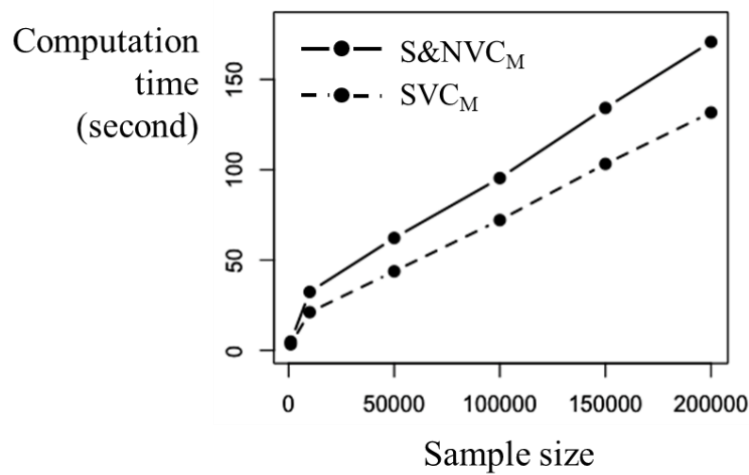

Figure 11: A comparison of computation times.

## An empirical application

This section applies S&NVC$_M$ to a land price analysis in the Ibaraki prefecture of Japan. The subsection below describes data and settings, and the next subsection explains modeling results.

### Outline

The response variables are the logarithms of officially assessed residential land prices in January 2015 (sample size: 647; see Figure 12). The covariates are distance to the nearest railway station [*Station_d*; km], and railway distance between the nearest station and Tokyo station [*Tokyo_d*; km], which is located about 30 km from the southwestern border of this prefecture, and anticipated flooding depth [*Flood*; km]. Their coefficients are denoted by $\beta_{Station}$, $\beta_{Tokyo}$, and $\beta_{Flood}$. This prefecture suffered from a major flood with 10,390 people being evacuated to shelters at its peak in October 2015; assessment of the impact from *Flood* is important in this area. All the variables are available from the National Land Numerical Information download service provided by the Ministry of Land, Infrastructure, Transport and Tourism (http://nlftp.mlit.go.jp/ksj-e/index.html).

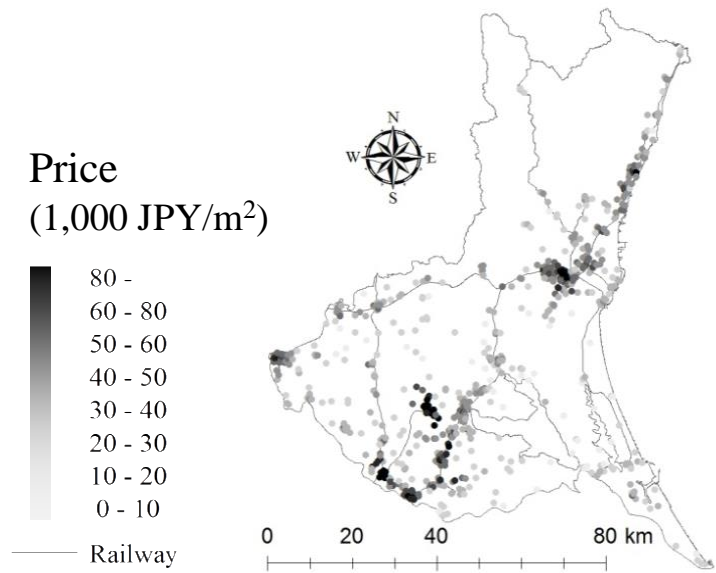

Figure 12: Residential land prices in Ibaraki prefecture in 2015.

Results

Table 1 summarizes the estimated share of SVC in each coefficient. The share is evaluated by $sd\left[\widehat{\boldsymbol{\beta}}_k^{(s)}\right]/(sd\left[\widehat{\boldsymbol{\beta}}_k^{(s)}\right] + sd\left[\widehat{\boldsymbol{\beta}}_k^{(n)}\right])$. We assume spatial variation only in the intercept following studies in spatial statistics. This table demonstrates that about half of the variation in $\boldsymbol{\beta}_{Station}$ and $\boldsymbol{\beta}_{Tokyo}$ is explained by SVC, whereas 98.2 % of the variation in

$\boldsymbol{\beta}_{Flood}$ is explained by SVC. Both spatial and non-spatial variation is present in the

coefficients; the share of SVC and NVC changes considerably depending upon $\boldsymbol{\beta}_k$.

Table 1: Estimated share of SVC, with spatial variation in only the intercept.

| | Intercept | Tokyo_d | Station_d | Flood |
|---|---|---|---|---|
| Share of SVC | 1.000 | 0.530 | 0.432 | 0.982 |

Figures 13 and 14 plot the predicted SVC $\boldsymbol{\beta}_k^{(s)}$ and NVC $\boldsymbol{\beta}_k^{(n)}$. For SVC, influence from Tokyo_d is weaker for nearby central cities, including Mito, Tsukuba, and Hitachi, probably because these cities are locally more influential than Tokyo. The predicted NVC for Tokyo_d becomes small at distances approaching those to the locations of the three central cities. Local subcenters appear to weaken influence from Tokyo. $\boldsymbol{\beta}_{Station}^{(s)}$ has a stronger negative impact near railways. This finding is intuitively reasonable. $\boldsymbol{\beta}_{Station}^{(n)}$ increases its negative value as the distance from the nearest station increases, taking its maximum negative value when the distance is around 2 km, whereas this impact declines as the distance increases beyond 2 km. 2 km might be a critical distance determining influence from a railway station. The SVC on Flood has large negative values around Mito city, which is the prefectural capital. Based on this finding, land prices in flood prone areas are appropriately discounted in Mito. In other words, the urban form is adaptive to flood risk in terms of residential land price.

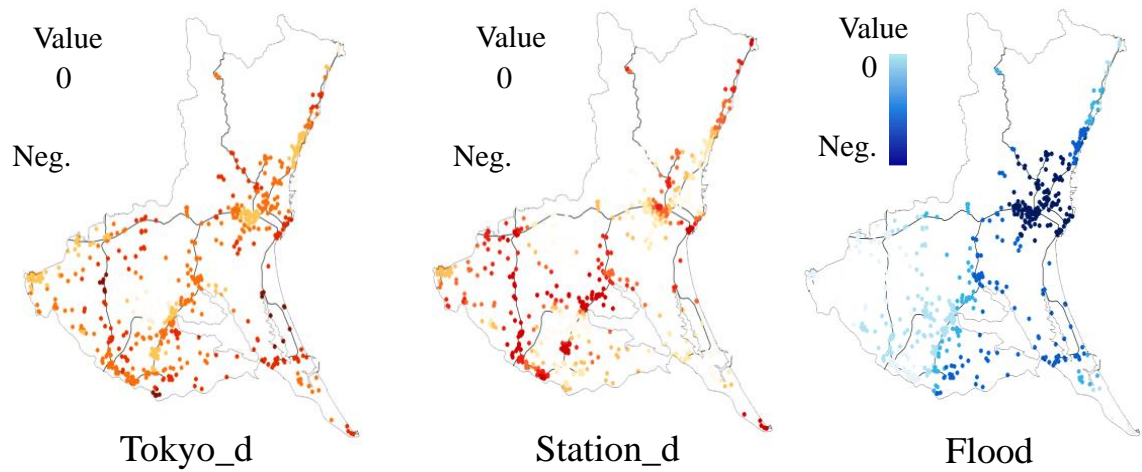

Figure 13: Estimated SVC. Lines denote the railway network.

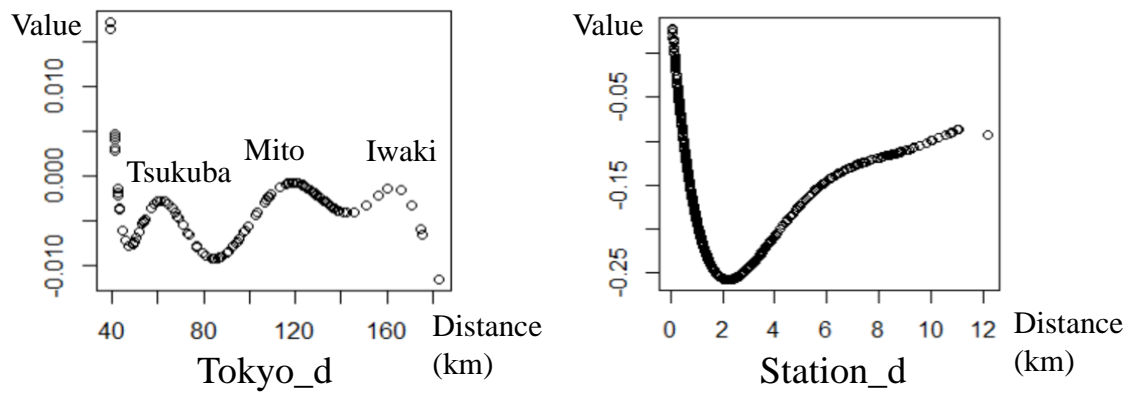

Figure 14: Estimated NVCs.

Finally, Figure 15 portrays predicted S&NVC. Although their map patterns are similar to those for SVC, the tendency of declining influence from Tokyo_d on nearby central cities, and the tendency of an increasing impact of Station_d along the railway, are clearer than for SVC. Mapping S&NVC appears to be useful for understanding varying relationships between covariates and response variables.

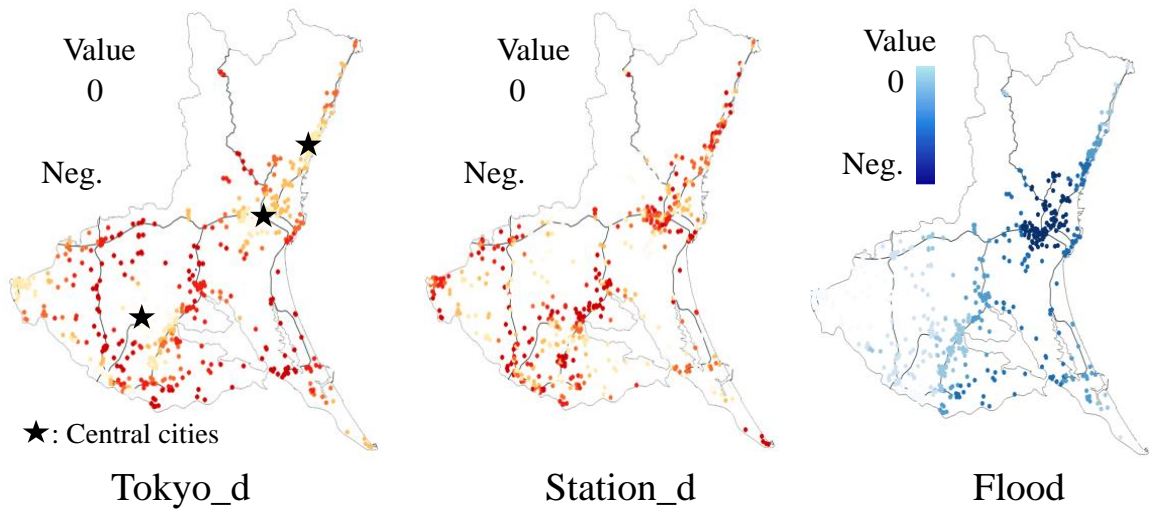

Figure 15: Estimated S&NVC. Lines denote the railway network.

## Concluding remarks

This study develops a Moran-eigenvector approach for predicting S&NVC that comprises computationally efficient SVC and NVC predictors. A Monte Carlo experiment suggests accuracy and stability of our approach, whereas an empirical land price analysis verifies the existence of spatial and non-spatial variation in regression coefficients. We also found that S&NVC modeling furnishes a remedy for spurious correlation or multicollinearity among coefficients, which is one of the biggest problems in SVC modeling. Although non-spatial aspects have been somewhat overlooked in spatial statistics, they might actually be a key to stable estimation of spatial models in the presence of spatially dependent covariates, which is a source of spurious correlation.

Our model is a particular type of spatial LMM that is readily extended to accommodate group effects, temporal effects, and other smooth effects (see Murakami and Griffith 2020). Spatio-temporal (ST) extension are especially important. LMM has actively been extended to ST models describing ST process in a 3-dimensional space using spatial, temporal, and ST basis functions (e.g., Kneib and Fahrmeir 2006; Griffith 2012; Augustin et al. 2013). Unfortunately, ST modeling requires a larger number of basis functions than spatial modeling. Sparse modeling might be useful to extend our approach to balance ST and non-ST variation. Seya et al. (2015) and Donegan (2020) perform Moran eigenvector-based sparse modeling. Extensions to dynamic ST modeling also pose an interesting task to predict SVC and NVC dynamically varying over time. Finley et al. (2012) and Baker et al. (2016) study dynamic SVC models.

Although we assume S&NVC = SVC + NVC, alternative specifications exist. For example, S&NVC may be defined by the product or product-sum of SVC and NVC. Furthermore, NVC in a $k$-th S&NVC can be defined using basis functions generated from $\mathbf{x}_{k'}$, where $k' \neq k$, or those generated from $\{\mathbf{x}_1, \dots, \mathbf{x}_K\}$. The SVC definition also can use non-Euclidean distance, including road network distance, or minimum cost distance (see Lu et al. 2017; 2018). A method for selecting an appropriate S&NVC specification is needed.

The S&NVC model will be implemented in the R package "spmoran" (https://cran.r-project.org/web/packages/spmoran/index.html).

## Appendix 1: Derivation of the SVC model (Eq. 3)

In geostatistics, SVC is defined by a mean-centered GP (see Gelfand et al., 2003):

$$\boldsymbol{\beta}_k^{(s)} = \mathbf{M}\ddot{\boldsymbol{\beta}}_k^{(s)}, \quad \ddot{\boldsymbol{\beta}}_k^{(s)} \sim N(\mathbf{0}, \tau_{k(s)}^2(\mathbf{C} + \mathbf{I})). \tag{A1}$$

where $\ddot{\boldsymbol{\beta}}_k^{(s)}$ is the GP before centering. Remember that the diagonals of the $\mathbf{C}$ matrix are assumed zeros. It is immediate to see that Eq. (A1) has the following expression:

$$\boldsymbol{\beta}_k^{(s)} \sim N(\mathbf{0}, \tau_{k(s)}^2\mathbf{M}(\mathbf{C} + \mathbf{I})\mathbf{M}). \tag{A2}$$

Suppose that $\mathbf{MCM} = \begin{bmatrix} \mathbf{E}^{(s)}, \mathbf{E}^{(-s)} \end{bmatrix} \begin{bmatrix} \boldsymbol{\Lambda} & \\ & \boldsymbol{\Lambda}^- \end{bmatrix} \begin{bmatrix} \mathbf{E}^{(s)\prime} \\ \mathbf{E}^{(-s)\prime} \end{bmatrix}$. Then, the $\mathbf{M}(\mathbf{C} + \mathbf{I})\mathbf{M}$ matrix can be expanded while referring to Murakami and Griffith (2015) as follows

$$
\begin{aligned}
\mathbf{M}(\mathbf{C} + \mathbf{I})\mathbf{M} &= \begin{bmatrix} \mathbf{E}^{(s)}, \mathbf{E}^{(-s)} \end{bmatrix} \begin{bmatrix} \boldsymbol{\Lambda} + \mathbf{I} & \\ & \boldsymbol{\Lambda}^- + \mathbf{I} - \mathbf{I}(\lambda_l = 0) \end{bmatrix} \begin{bmatrix} \mathbf{E}^{(s)\prime} \\ \mathbf{E}^{(-s)\prime} \end{bmatrix}, \\
&= \mathbf{E}^{(s)}(\boldsymbol{\Lambda} + \mathbf{I})\mathbf{E}^{(s)\prime} + \mathbf{E}^{(-s)}(\boldsymbol{\Lambda}^- + \mathbf{I} - \mathbf{I}(\lambda_l = 0))\mathbf{E}^{(-s)\prime}, \\
&= \mathbf{E}^{(s)}\boldsymbol{\Lambda}\mathbf{E}^{(s)\prime} + \mathbf{E}^{(-s)}\boldsymbol{\Lambda}^-\mathbf{E}^{(-s)\prime} + \begin{bmatrix} \mathbf{E}^{(s)}, \mathbf{E}^{(-s)} \end{bmatrix} \begin{bmatrix} \mathbf{E}^{(s)\prime} \\ \mathbf{E}^{(-s)\prime} \end{bmatrix} \\
&\quad + \begin{bmatrix} \mathbf{E}^{(s)}, \mathbf{E}^{(-s)} \end{bmatrix} \mathbf{I}(\lambda_l = 0) \begin{bmatrix} \mathbf{E}^{(s)\prime} \\ \mathbf{E}^{(-s)\prime} \end{bmatrix}, \\
&= \mathbf{E}^{(s)}\boldsymbol{\Lambda}\mathbf{E}^{(s)\prime} + \mathbf{E}^{(-s)}\boldsymbol{\Lambda}^-\mathbf{E}^{(-s)\prime} + \mathbf{I} - \mathbf{e}_N^{(s)}\mathbf{e}_N^{(s)\prime},
\end{aligned}
\tag{A3}
$$

where $\mathbf{I}(\lambda_l = 0)$ is a diagonal matrix whose $l$-th entry is 1 if $\lambda_l = 0$, and 0 otherwise.

Because $\mathbf{M}(\mathbf{C}+\mathbf{I})\mathbf{M} = \mathbf{MCM}+\mathbf{M}$ ($\mathbf{M}$ is an idempotent matrix) and $\mathbf{E}^{(s)}\boldsymbol{\Lambda}\mathbf{E}^{(s)'} + \mathbf{E}^{(-s)}\boldsymbol{\Lambda}^-\mathbf{E}^{(-s)'} = \mathbf{MCM}$, $\mathbf{I}-\mathbf{e}_N^{(s)}\mathbf{e}_N^{(s)'}$ in Eq. (A3) equals $\mathbf{MCM}+\mathbf{M}-\mathbf{MCM} = \mathbf{M} = \mathbf{I}-\mathbf{11}'/N$. In the end, we have

$$\mathbf{M}(\mathbf{C}+\mathbf{I})\mathbf{M} = \mathbf{E}^{(s)}\boldsymbol{\Lambda}\mathbf{E}^{(s)'} + \mathbf{E}^{(-s)}\boldsymbol{\Lambda}^-\mathbf{E}^{(-s)'} + \mathbf{I}-\mathbf{11}'/N. \qquad (A4)$$

Eq.(A4) suggests that the covariance matrix is factorized into the elements explaining positively dependent variation ( $\mathbf{E}^{(s)}\boldsymbol{\Lambda}\mathbf{E}^{(s)'}$ ), negatively dependent variation ($\mathbf{E}^{(-s)}\boldsymbol{\Lambda}^-\mathbf{E}^{(-s)'}$), independent variation ($\mathbf{I}$), and nuisance due to the centering ($-\mathbf{11}'/N$) that asymptotically disappears, respectively. In other words, $\mathbf{E}^{(s)}\boldsymbol{\Lambda}\mathbf{E}^{(s)'}$ is the only element explaining positive spatial dependence, which we assumed. After reducing the other elements, we obtain Eq. (A5) describing positively dependent SVC process:

$$\boldsymbol{\beta}_k^{(s)} = \mathbf{E}^{(s)}\boldsymbol{\gamma}_k^{(s)}, \qquad \boldsymbol{\gamma}_k^{(s)} \sim N\big(\mathbf{0}, \tau_{k(s)}^2\boldsymbol{\Lambda}\big), \qquad (A5)$$

Still, Eq.(A5) is restrictive in that it is incapable of modeling the spatial scale of $\boldsymbol{\beta}_k^{(s)}$. To overcome the limitation, we introduce the $\alpha_k$ parameter, which is equivalent to the smoothness parameter in geostatistics, as follows:

$$\boldsymbol{\beta}_k^{(s)} = \mathbf{E}^{(s)}\boldsymbol{\gamma}_k^{(s)}, \qquad \boldsymbol{\gamma}_k^{(s)} \sim N\big(\mathbf{0}, \tau_{k(s)}^2\boldsymbol{\Lambda}^{\alpha_k}\big), \qquad (A6)$$

which equals Eq.(3). As explained in Murakami and Griffith (2020), the expectation of the Moran coefficient of $\boldsymbol{\beta}_k^{(s)}$ converges 0 meaning random spatial distribution as $\alpha_k \to -\infty$ whereas converges to the theoretical maximum, meaning the strongest positive spatial dependence measured by the Moran coefficient, as $\alpha_k \to \infty$ (See Figure 3). We apply Eq.(A6) for the SVC modeling.

## Appendix.2. The fast REML

Our S&NVC model has the following linear mixed effects model (LMM) representation:

$$\mathbf{y} = \mathbf{X}\mathbf{b} + \mathbf{E}_{1:K}\mathbf{V}_{1:K}(\boldsymbol{\theta}_{1:K})\mathbf{u}_{1:K} + \boldsymbol{\varepsilon}, \quad \mathbf{u}_{1:K} \sim N(\mathbf{0}, \sigma^2\mathbf{I}), \quad \boldsymbol{\varepsilon} \sim N(\mathbf{0}, \sigma^2\mathbf{I}). \qquad \text{(A7)}$$

$\mathbf{X} = [\mathbf{x}_1, \cdots, \mathbf{x}_K]$, $\mathbf{b} = [b_1, \dots, b_K]'$, $\mathbf{E}_{1:K} = [(\mathbf{x}_1 \circ \mathbf{E}^{(s)}), \dots, (\mathbf{x}_K \circ \mathbf{E}^{(s)}), (\mathbf{x}_1 \circ \mathbf{E}_1^{(n)}), \dots, (\mathbf{x}_K \circ \mathbf{E}_K^{(n)})]$, and $\mathbf{u}_{1:K} = [\mathbf{u}_1^{(s)\prime}, \dots, \mathbf{u}_K^{(s)\prime}, \mathbf{u}_1^{(n)\prime}, \dots, \mathbf{u}_K^{(n)\prime}]'$. $\mathbf{V}_{1:K}(\boldsymbol{\theta}_{1:K})$ is a block diagonal matrix whose $k$-th block is $\frac{\tau_{k(s)}}{\sigma}\boldsymbol{\Lambda}^{\alpha_k/2}$ and $(K + k)$-th block is $\frac{\tau_{k(n)}}{\sigma}\mathbf{I}$. The block diagonal matrix implicitly assumes independence across the SVCs. This assumption considerably improves computational efficiency while maintaining the SVC modeling accuracy, as shown in Murakami and Griffith (2019b). $\boldsymbol{\theta}_{1:K} = \{\boldsymbol{\theta}_1, \dots, \boldsymbol{\theta}_K\}$ summarizes variance parameters, where $\boldsymbol{\theta}_k \in \{\tau_{k(s)}^2, \tau_{k(n)}^2, \alpha_k\}$. Eq. (A7) implies

$$\begin{aligned}
\boldsymbol{\beta}_k^{(s)} &= \frac{\tau_{k(s)}}{\sigma}\mathbf{E}^{(s)}\boldsymbol{\Lambda}^{\alpha_k/2}\mathbf{u}_k^{(s)}, \quad \mathbf{u}_k^{(s)} \sim N(\mathbf{0}, \sigma^2\mathbf{I}), \quad \text{and} \\
\boldsymbol{\beta}_k^{(n)} &= \frac{\tau_{k(n)}}{\sigma}\mathbf{E}^{(n)}\mathbf{u}_k^{(n)}, \quad \mathbf{u}_k^{(n)} \sim N(\mathbf{0}, \sigma^2\mathbf{I}).
\end{aligned} \qquad \text{(A8)}$$

We estimate Eq.(A7) using the fast REML method of Murakami and Griffith

(2019b), which was developed to estimate LMM that includes Eq. (A7) as a special case.

The restricted log-likelihood for our model yields

$$l_R(\boldsymbol{\theta}_{1;K}) = -\frac{1}{2} ln \begin{vmatrix} \mathbf{X'X} & \mathbf{X'E}_{1:K}\mathbf{V}_{1;K}(\boldsymbol{\theta}_{1;K}) \\ \mathbf{V}_{1;K}(\boldsymbol{\theta}_{1;K})\mathbf{E'}_{1:K}\mathbf{X} & \mathbf{V}_{1;K}(\boldsymbol{\theta}_{1;K})\mathbf{E'}_{1:K}\mathbf{E}_{1:K}\mathbf{V}_{1;K}(\boldsymbol{\theta}_{1;K}) + \mathbf{I} \end{vmatrix}$$
$$- \frac{N-K}{2}\left(1 + ln\left(2\pi \frac{\left\|\mathbf{y} - \mathbf{X\hat{b}} - \mathbf{E}_{1:K}\mathbf{V}_{1:K}(\boldsymbol{\theta}_{1;K})\mathbf{\hat{u}}_{1:K}\right\|^2 + \|\mathbf{\hat{u}}_{1:K}\|^2}{N-K}\right)\right), \quad \text{(A9)}$$

where

$$\begin{bmatrix} \mathbf{\hat{b}} \\ \mathbf{\hat{u}}_{1;K} \end{bmatrix} =$$
$$\begin{bmatrix} \mathbf{X'X} & \mathbf{X'E}_{1:K}\mathbf{V}_{1;K}(\boldsymbol{\theta}_{1;K}) \\ \mathbf{V}_{1;K}(\boldsymbol{\theta}_{1;K})\mathbf{E'}_{1:K}\mathbf{X} & \mathbf{V}_{1;K}(\boldsymbol{\theta}_{1;K})\mathbf{E'}_{1:K}\mathbf{E}_{1:K}\mathbf{V}_{1;K}(\boldsymbol{\theta}_{1;K}) + \mathbf{I} \end{bmatrix}^{-1} \begin{bmatrix} \mathbf{X'y} \\ \mathbf{V}_{1;K}(\boldsymbol{\theta}_{1;K})\mathbf{E'}_{1:K}\mathbf{y} \end{bmatrix}. \quad \text{(A10)}$$

The REML balances the residual variance $\left\|\mathbf{y} - \mathbf{X\hat{b}} - \mathbf{E}_{1:K}\mathbf{V}_{1:K}(\boldsymbol{\theta}_{1;K})\mathbf{\hat{u}}_{1:K}\right\|^2$, which

quantifies model accuracy, and the variance of random coefficients $\mathbf{\hat{u}}_{1:K}$, which quantifies model complexity, generating SVC and NVC (see Eq. A2). The variance estimator $\mathbf{\hat{\theta}}_{1;K}$ maximizing $l_R(\boldsymbol{\theta}_{1;K})$ is identified computationally quite efficiently; see

Murakami and Griffith (2019b, 2020) for further detail.

Eq. (A10) provides the best linear unbiased estimator (BLUE) for $\mathbf{b}$ and the

best linear unbiased predictor (BLUP) for $\mathbf{u}_{1;K}$ (Henderson, 1975; Searle, 1997). Given

$\mathbf{\hat{\theta}}_{1;K}$, coefficients $[\mathbf{\hat{b}'}, \mathbf{\hat{u}'}_{1;K}]'$ are obtained using Eq. (A10). After that, the S&NVC

predictor, which is unbiased and minimizes the expected predictive error variance, is

obtained as follows (see Eqs. 1 and A8):

$$\widehat{\boldsymbol{\beta}}_k = \hat{b}_k \mathbf{1} + \widehat{\boldsymbol{\beta}}_k^{(s)} + \widehat{\boldsymbol{\beta}}_k^{(n)},$$

$$\widehat{\boldsymbol{\beta}}_k^{(s)} = \frac{\hat{\tau}_{k(s)}}{\hat{\sigma}} \mathbf{E}^{(s)} \boldsymbol{\Lambda}^{\hat{\alpha}_k/2} \widehat{\mathbf{u}}_k^{(s)}, \qquad \widehat{\boldsymbol{\beta}}_k^{(n)} = \frac{\hat{\tau}_{k(n)}}{\hat{\sigma}} \mathbf{E}^{(n)} \widehat{\mathbf{u}}_k^{(n)}. \tag{A11}$$

$\hat{\tau}_{k(s)}$ and $\hat{\tau}_{k(n)}$ estimate the variances of $\widehat{\boldsymbol{\beta}}_k^{(s)}$ and $\widehat{\boldsymbol{\beta}}_k^{(n)}$ respectively. If both $\hat{\tau}_{k(s)}$ and $\hat{\tau}_{k(n)}$ are zeros, the S&NVC predictor $\widehat{\boldsymbol{\beta}}_k$ becomes a constant $\hat{b}_k \mathbf{1}$. In addition, the variance of the coefficients is deflated if data is noisy (i.e., large $\hat{\sigma}$). Thus, our estimation balances noise, spatial, and non-spatial variation in the coefficients.